\documentclass[aps,preprintnumbers,showpacs,showkeys]{revtex4}
\usepackage{epsfig}
\usepackage{amssymb,amsmath,amsfonts,amsthm,graphicx}
\graphicspath{{./Figures/}}             

\usepackage{multirow}

\newcommand{\beq}{\begin{equation}}
\newcommand{\eeq}{\end{equation}}
\newcommand{\bea}{\begin{eqnarray}}
\newcommand{\eea}{\end{eqnarray}}

\newcommand{\eq}{\begin{equation}}
\newcommand{\en}{\end{equation}}
\newcommand{\eqa}{\begin{eqnarray}}
\newcommand{\ena}{\end{eqnarray}}
\newcommand{\half}{\frac{1}{2}}
\newcommand{\tr}{\mbox{Tr}}

\begin{document}

\author{V.~G.~Bornyakov}
\affiliation{NRC ``Kurchatov Institute'' - IHEP,
 Protvino, 142281 Russia, \\
NRC ``Kurchatov Institute'' - ITEP, Moscow, 117218 Russia }

\author{I.~Kudrov}
\affiliation{NRC ``Kurchatov Institute'' - ITEP, Moscow, 117218 Russia }

\author{R.~N.~Rogalyov}
\affiliation{NRC "Kurchatov Institute" - IHEP, 142281 Protvino, Russia}



\begin{abstract}
We study decomposition of $SU(2)$ gauge field into monopole and monopoleless components. 
After fixing the Maximal Abelian gauge in $SU(2)$ lattice gauge theory with Wilson action
we decompose the nonabelian gauge field
into the Abelian field created by monopoles and the modified nonabelian field with monopoles
removed. We then calculate respective static potentials in the fundamental and adjoint
representations and confirm earlier findings that the sum of these potentials approximates 
the nonabelian static potential with good precision at all distances considered. 
Repeating these computations at three lattice spacings we find that in both representations the approximation becomes better with decreasing lattice spacing. Our results thus suggest that this approximation becomes exact in the continuum limit.
We further find the same relation (for one lattice spacing)  to be valid also in the cases of improved lattice action and in the theory with quarks. 

\end{abstract}

\title{Decomposition of the SU(2) gauge field in the Maximal Abelian gauge}



\maketitle

\section{Introduction}
\label{section1}
We study numerically the lattice $SU(2)$ 
gluodynamics in the Maximal Abelian gauge (MAG) 
and consider decomposition of the lattice gauge 
field $U_\mu(x)$ 
\begin{equation}
 U_\mu(x) =  U_\mu^{mod}(x) U_\mu^{mon}(x)  
\label{eq:decomposition}
\end{equation}
where $U_\mu^{mon}(x)$ is the monopole component 
and $U_\mu^{mod}(x)$ is respectively the monopoleless 
component which we also will call a modified gauge field. 
By modification we understand removal of monopoles.

It is well known \cite{suzuki1,suzuki2,bbms,bm,Sakumichi:2014xpa} that 
after performing the Abelian projection in the MAG 
\cite{Kronfeld:1987ri,thooft2}, the Abelian 
string tension calculated from the Abelian static potential 
is very close to the nonabelian string tension and 
the corresponding coefficient of the Coulomb term is about 1/3 of 
that in the nonabelian static potential. The former observation, 
like many others, supports the concept of Abelian dominance 
(for a review see e.g. \cite{review}).
It was further discovered \cite{suzuki3,stack,bbms} that the monopole static 
potential also has string tension close to the nonabelian one and small 
coefficient of the Coulomb term.  These observations are in agreement with 
conjecture that monopole degrees of 
freedom are responsible for confinement \cite{thooft}. 
It is then interesting to see what kind of static
potential one obtains if the monopole contribution into the 
gauge field switches off, that is, if only off-diagonal 
gluons and the so called photon part of the Abelian gluon 
field are left interacting with static quarks. 

Previously computations of this kind were made in 
\cite{miyamura,Kitahara:1998sj}, where it 
was shown that the topological charge, chiral condensate and
effects of chiral symmetry breaking in quenched light hadron spectrum
disappear after 
removal of the monopole contribution from the relevant operators. Similar 
computations were made within the scope of the $Z_2$ projection studies 
\cite{deforcrand}. It was shown that modified gauge field with removed 
projected center vortices (P-vortices) produces Wilson loops without area law, 
i.e. devoid of the confinement property. We  do a similar 
removal with monopoles. 
We consider three types of the static potential: $V_{mod}(r)$ obtained from the Wilson 
loops of the modified gauge field $U_\mu^{mod}(x)$, 
$V_{mon}(r)$ obtained from the Wilson loops of the monopole gauge field $U_\mu^{mon}(x)$ and the sum of these two static potentials. 
 
The decomposition (\ref{eq:decomposition}) was first considered in \cite{Bornyakov:2005hf}. 
It was demonstrated for one value of the lattice spacing that $V_{mod}(r)$ could be well fitted by purely Coulomb fit function and the sum $V_{mod}(r)+V_{mon}(r)$ 
was a good approximation of the original nonabelian static potential, 
$ V(r)$, at all distances. 

Here we study this phenomenon at three lattice spacings 
using the Wilson lattice gauge field action and thus 
we can make conclusions about the continuum limit. 
We also present the results for one lattice spacing 
obtained with the improved lattice field action thus checking the universality. 
Furthermore, we present results for the $SU(2)$ theory with dynamical quarks, i.e. for QC$_2$D. 

The paper is organized as follows. In the next section we introduce relevant 
definitions and describe details of our computations. In section 3 
results for the static potential are presented. Section 4 is devoted to 
discussion and conclusions.  

\section{Definitions and simulation details}

We consider the $SU(2)$ lattice gauge theory after fixing MAG. The Abelian projection means coset 
decomposition of the nonabelian lattice gauge field $U_\mu(x)$ into the Abelian field $u_\mu(x)$ and the coset field $C_\mu(x)$ \footnote{The necessary derivations was presented
in \cite{Bornyakov:2005hf}. Here we briefly repeat them for the reader's convenience.}:
\eq
U_\mu(x) = C_\mu(x) u_\mu(x)\quad ,
\label{coset}
\en
The Abelian gauge field can be further decomposed into the  monopole (singular)
part $u^{mon}_\mu(x)$ and the photon (regular) part $u^{ph}_\mu(x)$ 
\cite{svs}:
\eq
u_\mu(x) = u^{mon}_\mu(x) u^{ph}_\mu(x)\ .
\en
In terms of the corresponding angles it has the form
\eq
\theta_\mu(x) = \theta^{mon}_\mu(x) + \theta^{ph}_\mu(x)\,,
\en
where $\theta_\mu(x) \in (-\pi,\pi]$  is defined by  
$u_\mu(x)= e^{i \theta_\mu(x)}$, and $\theta^{mon,ph}_\mu(x)$ are defined analogously. 
$\theta^{mon}_\mu(x)$ can be presented as follows:
\eq
\theta^{mon}_\mu(x)  = -2 \pi \sum_{y} D(x-y) \partial_{\nu}^{'}
m_{\nu\mu}(y) \,,
\label{factor3}
\en
where $D(x)$ is lattice inverse Laplacian, $\partial_{\nu}^{'}$ is lattice backward derivative, $m_{\nu\mu}(x)$ are Dirac plaquettes. 
This solution satisfies the Lorenz gauge condition 
$\partial'_{\mu} \theta_{mon}(s,\mu) = 0$.
We calculate the usual Wilson loops
\eq
W(C) = \half \tr {\cal{W}}(C)\,,\quad      {\cal{W}}(C) = \left(\prod_{l \in C} U(l)\right) \quad,
\label{wnonab}
\en
the monopole Wilson loops
\eq
W_{mon}(C) = \half \tr\left(\prod_{l \in C} u^{mon}(l)\right) \quad,
\label{wmon}
\en
and the nonabelian Wilson loops with removed monopole contribution 
\eq
W_{mod}(C) = \half \tr  {\cal{W}}_{mod}(C)\,,\quad   {\cal{W}}_{mod}(C)  =   \left(\prod_{l \in C} \tilde{U}(l)\right) \quad,
  \label{woff}
\en
where the modified nonabelian gauge field $\tilde{U}_\mu(x)$ is defined as 
\eq
\tilde{U}_\mu(x)= C_\mu(x)u^{ph}_\mu(x).
\en
Note that $u^{ph}(x)$ is the Abelian projection
of $\tilde{U}_\mu(x)$ and involves no monopoles.

It is known that MAG fixing leaves $U(1)$ gauge symmetry unbroken. 
The general form of the $U(1)$ gauge transformation 
is given by
\eq
\theta^\prime_\mu(x)  = \theta_\mu(x)  +  \partial_{\mu} \omega(x)  + 2\pi n_\mu(x)\,,
\en 
where 
$\theta^\prime_\mu(x), \omega(x) \in (-\pi,\pi]$,  $ n_\mu(x) = 0, \pm 1$. 
Thus there are 'small' gauge transformations with $ n_\mu(x)=0$ and 
'large' gauge transformations with $ n(s,\mu)=\pm 1$.
The monopole Wilson loop $ W_{mon}(C)$ is invariant under these 
gauge transformations.  This is not true for $ W_{mod}(C)$. 
It was shown in \cite{Bornyakov:2005hf} that $ W_{mod}(C)$ is invariant only under
'small' gauge transformations and it is necessary to remove 'large' gauge transformations.
To this end we fix the Landau $U(1)$ gauge using the gauge condition
\eq
\max_{\omega} \sum_{x,\mu} cos(\theta^\prime_\mu(x)\; .
\label{landau}
\en
Up to Gribov copies this conditions fixes configuration of Dirac plaquettes $m_{\mu\nu}(x)$ completely.
Fixing $U(1)$ Landau gauge is excessive for our purposes but is eligible for calculations of $W_{mod}$.

We calculated $r\times t$ rectangular Wilson loops $W(r,t)$, $W_{mon}(r,t)$ and $W_{mod}(r,t)$.
To extract respective  static potentials the APE smearing \cite{ape}
has been employed. 
Computations were done with the Wilson lattice action at $\beta=2.4, 2.5$ 
on $24^4$ lattices and at $\beta=2.6$ on $32^4$ lattices using 100 statistically independent
configurations. With the tadpole improved action the simulation were made at $\beta=3.4$ on $24^4$ lattices.
The simulations in QC$_2$D were made on $32^4$ lattice with small lattice spacing \cite{Bornyakov:2017txe}.
To fix MAG, the simulated annealing  algorithm \cite{bbms} with one gauge copy was used.

\section{Static potential in fundamental and adjoint representations}
\label{section2}

\begin{figure}[htb]
\hspace*{-0mm}
\includegraphics[width=8cm,angle=0]{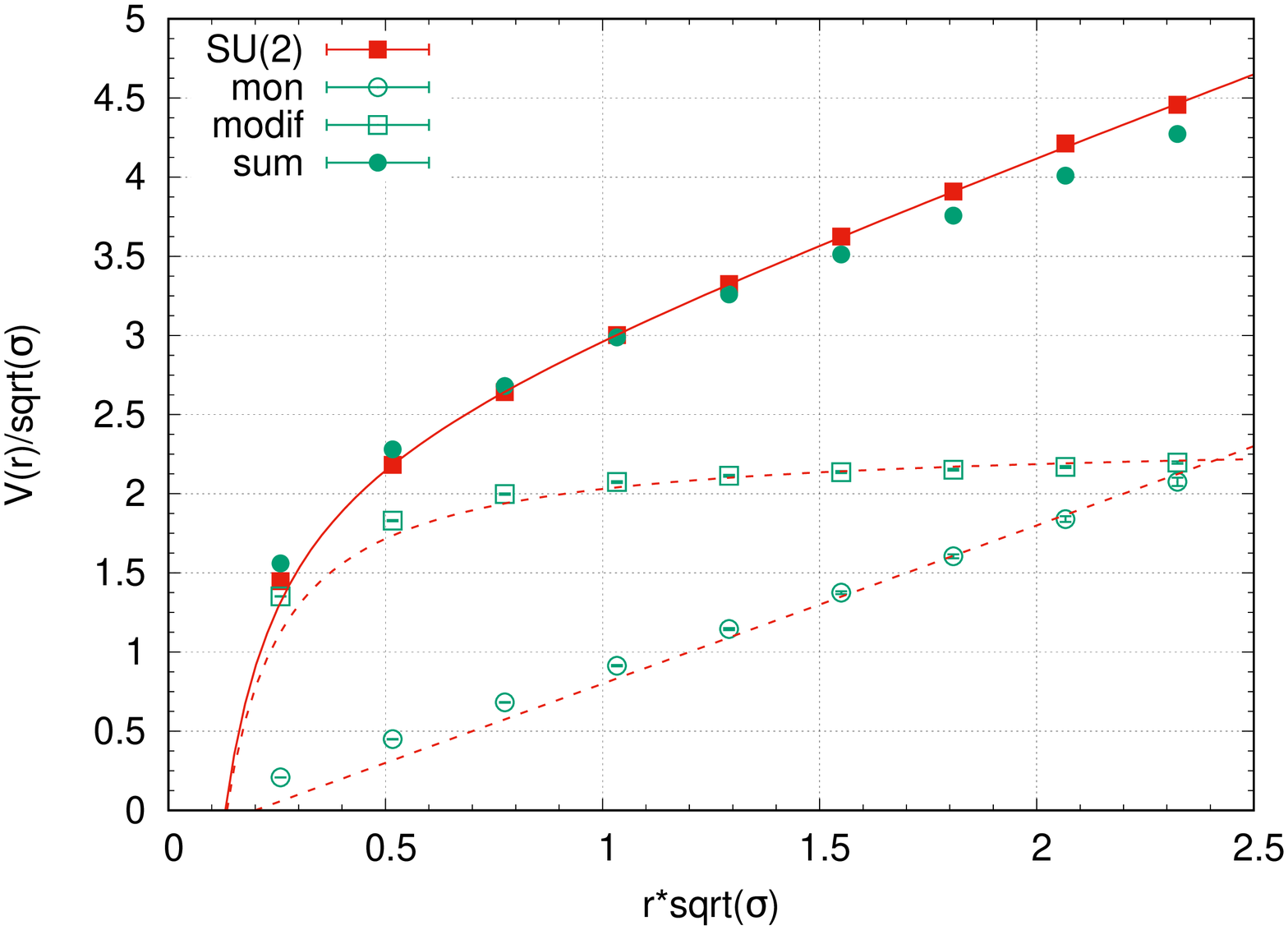}
\hspace*{5mm}
\includegraphics[width=8cm,angle=0]{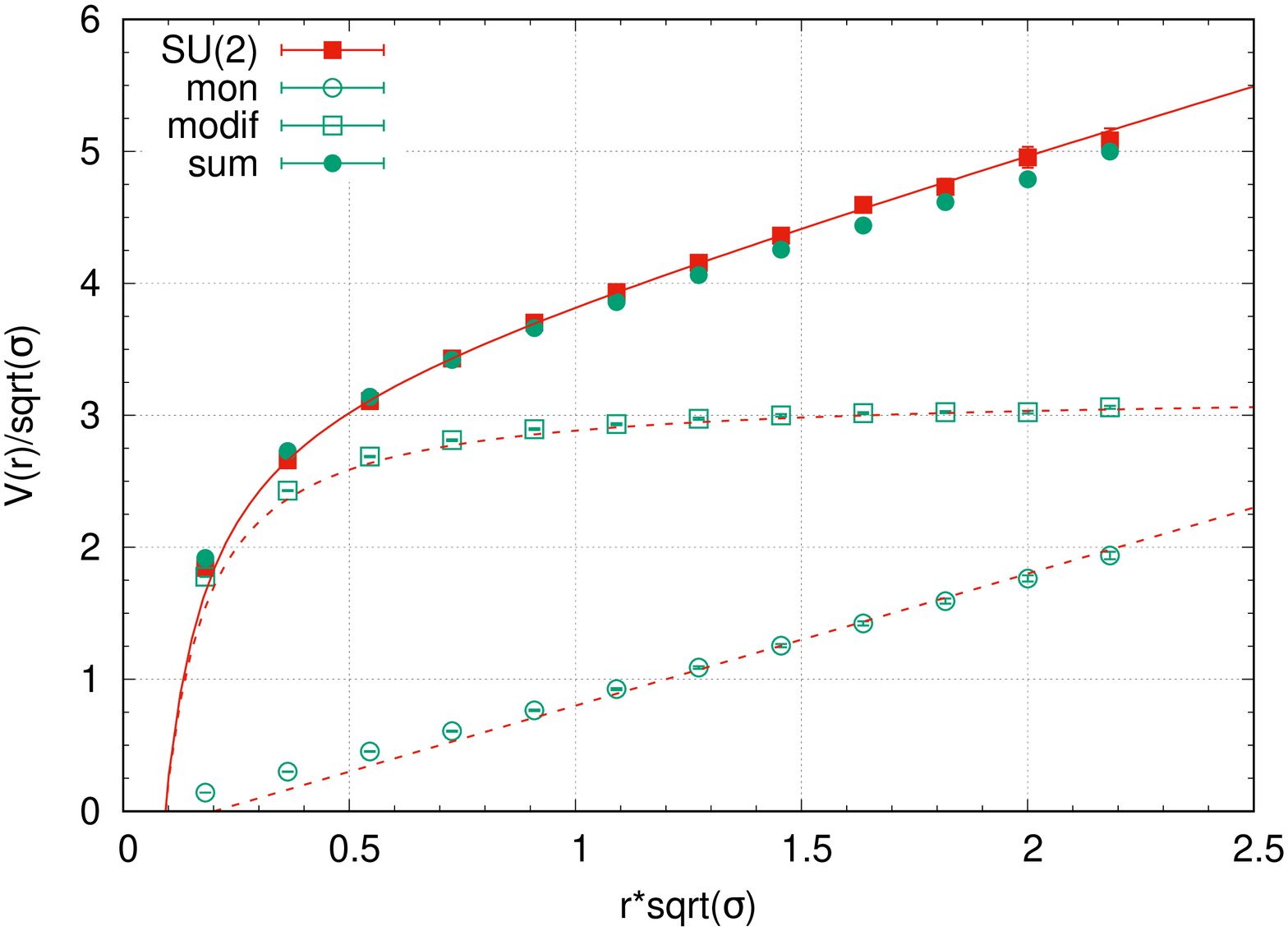}
\includegraphics[width=8cm,angle=0]{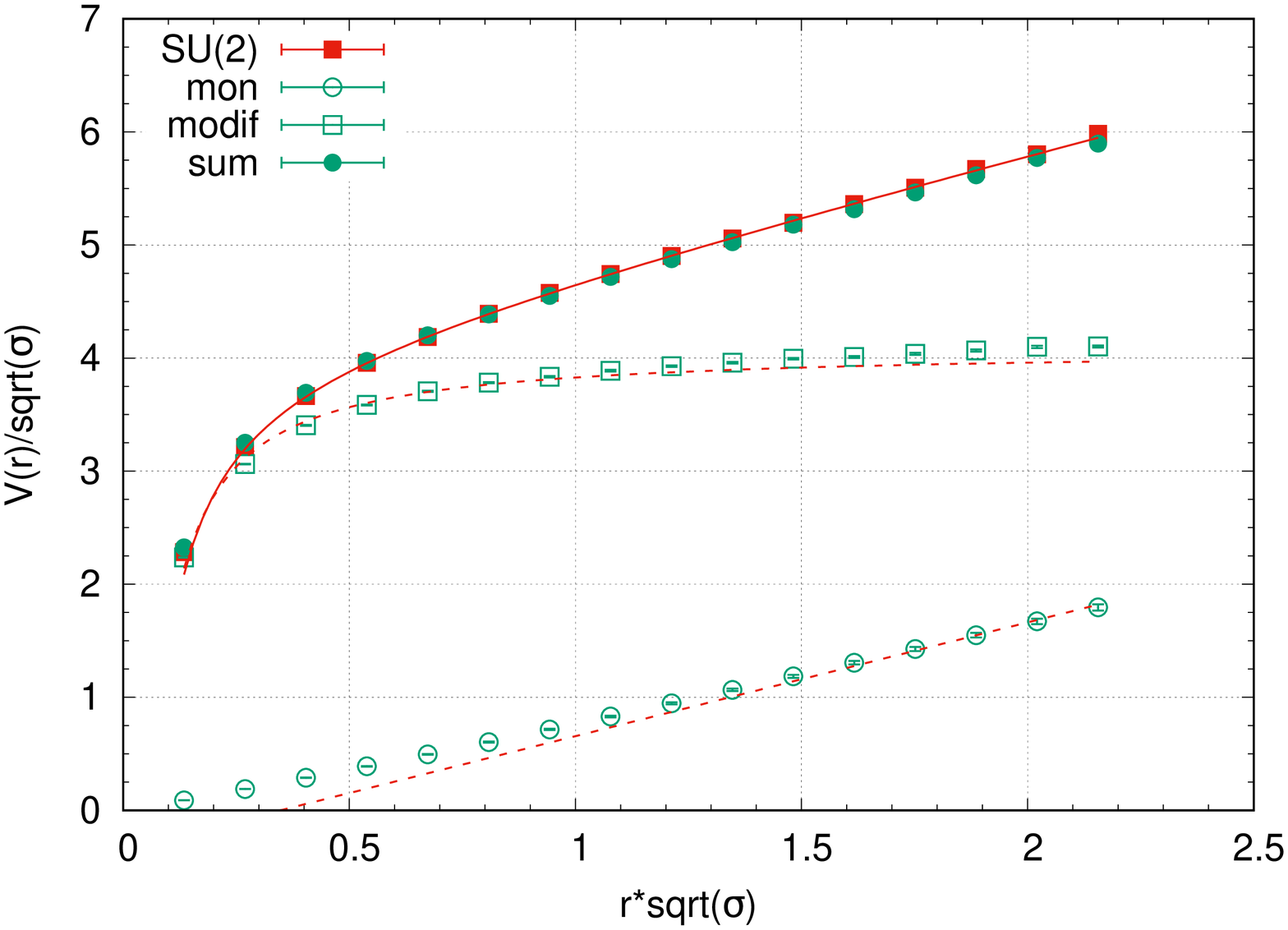}

\caption{Comparison of the nonabelian potential $V(r)$ (filled squares) with the 
sum  $V_{mod}(r)+V_{mon}(r)$ (filled circles) for $\beta=2.4$ (left upper panel), 
$\beta=2.5$ (right upper panel), $\beta=2.6$ (lower panel). 
$V_{mod}(r)$ (empty squares) and $V_{mon}(r)$ (empty circles) are also depicted. The solid curve 
shows the fit to  eq.~(\ref{eq:fit}).
Two dashed curves show its Coulomb and linear terms with adjusted constants. }
\label{potentials1}
\end{figure}

We present our results for the sum $ V_{mon}(r)+V_{mod}(r)$ and compare it with
the nonabelian potential $V(r)$ in Fig.\ref{potentials1} 
for lattice Wilson action and three lattice spacings. One can see that the nonabelian 
static potential $V(r)$ is well approximated by this sum, i.e.
\eq
V(r) \approx V_{mon}(r)+V_{mod}(r)\,.
\label{decomp}
\en
This observation can be formulated in the following way:
potential $V(r)$ between static sources interacting with  
the nonabelian gauge field $U_\mu(x)$ can be approximated
by the sum of the potential $V_{mon}(r)$ between the sources 
interacting only with the monopole field $U_\mu^{mon}(x)$ 
and the potential $V_{mod}(r)$ between the sources interacting 
only with the modified (monopoleless) field $U_\mu^{mod}(x)$. 

We fitted all static potentials to the fit function 
\eq
V(r)=V_0+\alpha/r+\sigma r.
\label{eq:fit}
\en
The results for the string tension $\sigma a^2$ and the Coulomb coefficient $\alpha$ are presented in Table~\ref{results}, irrelevant parameter $V_0$ is not shown.

\begin{figure}[htb]
\hspace*{-0mm}
\includegraphics[width=8cm,angle=0]{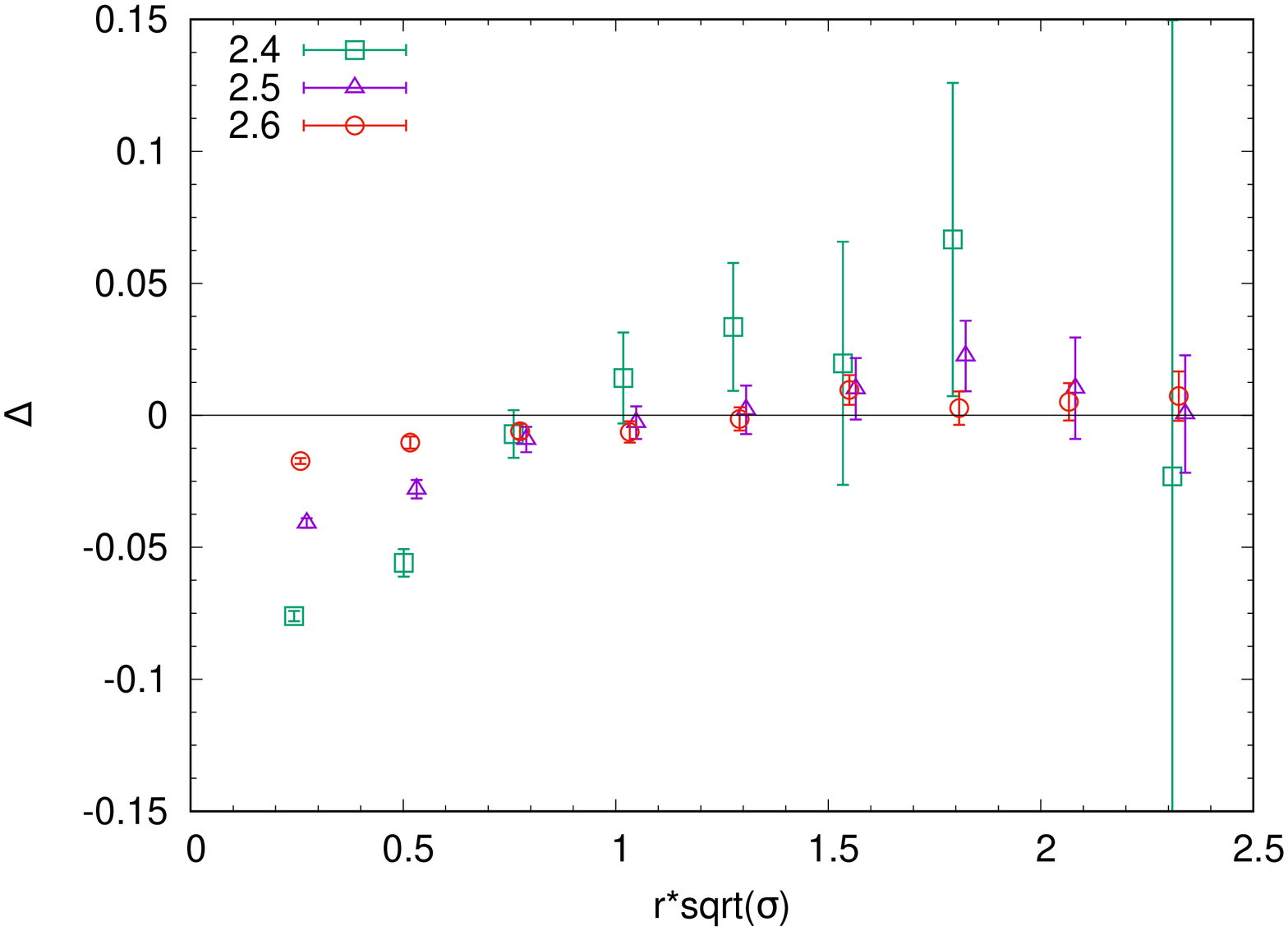}
\caption{The relative deviation $\Delta(r)$ defined in eq.~(\ref{eq:reldev}) vs. distance $r$ 
for three values 
of the 
coupling constant $\beta$.
}
\label{reldev}
\end{figure}

One can see that the agreement between  $V_{mon}(r)+V_{mod}(r)$ and $V(r)$ improves with decreasing lattice spacing. This is the main result of this paper. To make it more explicit 
we show in Fig.~\ref{reldev} the relative deviation determined as follows:
\beq
\Delta(r) = \frac{V(r) -  (V_{mon}(r)+V_{mod}(r))}{V(r)}\;.
\label{eq:reldev}
\eeq
More extended study  with increased precision and enlarged set of lattices is needed to make final conclusion about the continuum limit.

In Fig.\ref{potentials1} we also show the monopole $V_{mon}(r)$ and the modified 
field $V_{mod}(r)$ potentials separately.
We find that $V_{mon}(r)$ is linear at large distances and has small 
curvature at small distances, which can be well fitted by the Coulomb behavior with 
small positive coefficient.  
The slope of  $V_{mon}(r)$ at large distances
agrees better and better with that of $V(r)$ with decreasing lattice spacing.
We shall note that increasing of the ratio $\sigma_{mon}/\sigma$ with decreasing lattice spacing
was reported before in \cite{Bornyakov:2001ux}. 

It can be seen that $V_{mod}(r)$ is of Coulombic form. Indeed it can be very  well fitted by 
the fitting function
$V^{mod}_0 - \alpha_{mod}/r$ with $\alpha_{mod}=0.27(1)$ for $\beta=2.5$
and similar values for $\beta=2.4, 2.6$. One can see from Fig.\ref{potentials1} 
that $V_{mod}(r)$ is in a very good agreement with the Coulombic part of $V(r)$. 
Thus removing the monopole contribution 
from the Wilson loop operator leaves
Wilson loop which has no area law, i.e., the confinement property is lost. 
This result is similar to that obtained
in \cite{deforcrand} after removing P-vortices.



\begin{table}[tb]
\begin{center}
\caption{Parameters of the potentials obtained by fits to the function (\ref{eq:fit})
} 

\begin{tabular}{|c|c|c|c|c|c|c|}
\hline
{\multirow{2}{*}{Potential}} & \multicolumn{2}{c|}{$\beta=2.4$} & \multicolumn{2}{c|}{$\beta=2.5$} & \multicolumn{2}{c|}{$\beta=2.6$} \\
\cline{2-7}
 & $ ~~~~\sigma a^2 \qquad$  & $\alpha$ &$\sigma a^2$  & $\alpha$ & $\sigma a^2$ & $\alpha$  \\
\hline
$V$             & ~0.067(1) & ~-0.31(1)   & ~0.033(1) & ~-0.290(4) & ~0.0184(5) & ~-0.25(1) \\ 
$V_{mon}+V_{mod}$  & ~0.058(1)& ~-0.27(1) & ~0.030(1) & ~-0.27(1)  & ~0.0175(4)& ~-0.25(1) \\
$V_{mon}$          & ~0.060(1) & -0.002(1) & ~0.030(1) & 0.015(3) & ~0.0167(6) & 0.06(2)  \\
$V_{mod}$          & -          & ~-0.25(1) &  -       & ~-0.27(1) & 0.002(1)& ~-0.27(1)\\
\hline
\end{tabular}
\label{results}
\end{center}
\end{table}

\begin{figure}[htb]
\includegraphics[width=8cm,angle=0]{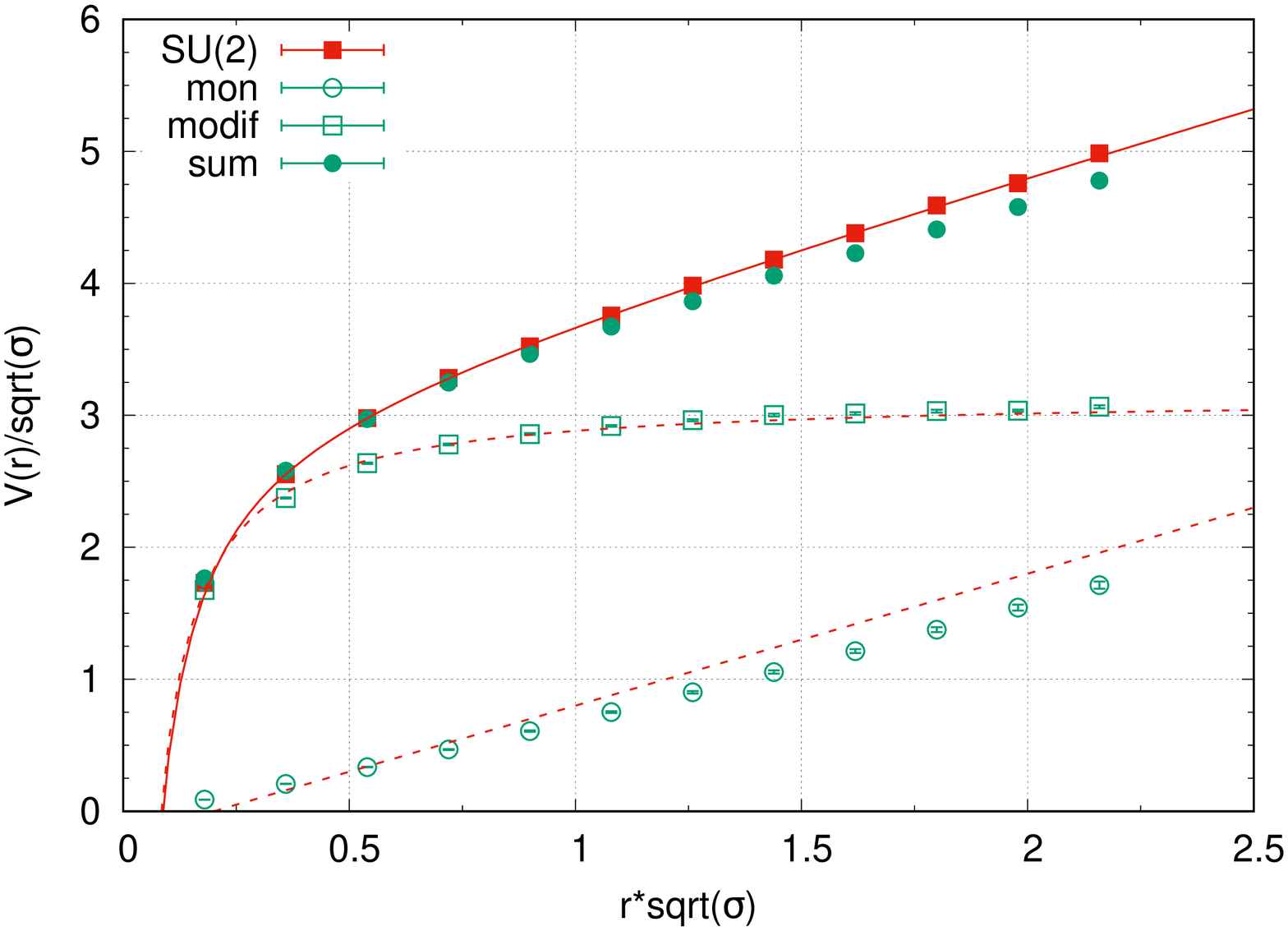}
\includegraphics[width=8cm,angle=0]{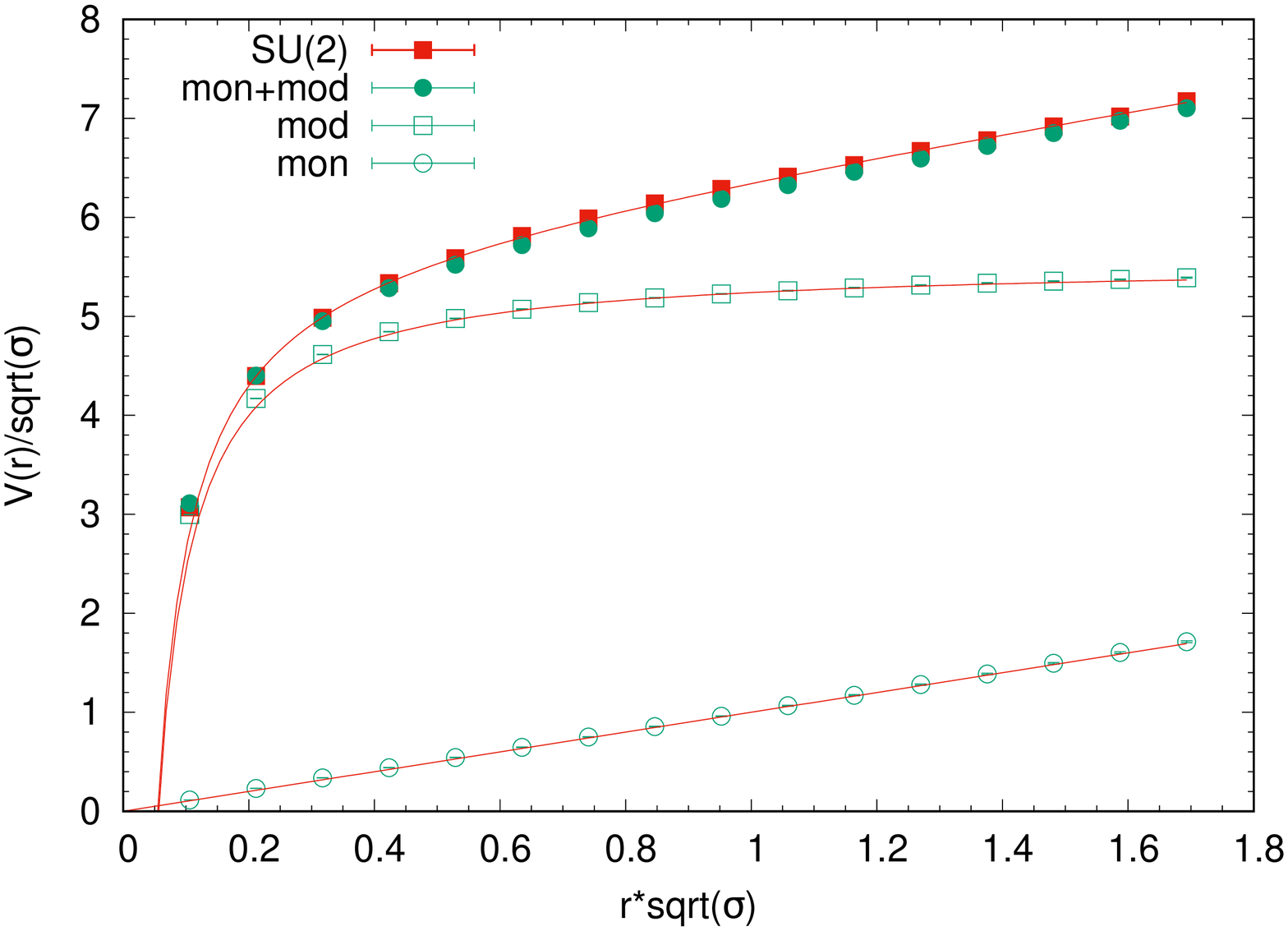}
\caption{Comparison of the nonabelian potential $V(r)$ (filled squares) with the 
sum  $V_{mod}(r)+V_{mon}(r)$ (filled circles) for improved action at $\beta=3.4$ (left) and 
for QC$_2$D (right). $V_{mod}(r)$ (empty squares) 
and $V_{mon}(r)$ (empty circles) are also shown. 
The solid curve and dashed curves carry same meaning as in Fig.~\ref{potentials1}. }
\label{potentials2}
\end{figure}

Apart from approach to the continuum limit we studied the question of universality of the decomposition eq.~(\ref{decomp}). The simulations were made with 
the tadpole improved action at $\beta=3.4$. 
The lattice spacing at this coupling 
is approximately equal to that of the Wilson action at $\beta=2.5$.
The result is presented in the Fig.~\ref{potentials2} (left). One can see that agreement between $V(r)$ and $V_{mod}(r)+V_{mon}(r)$ is nearly as good as in Fig.~\ref{potentials1} for $\beta=2.5$. 

\begin{figure}[htb]
\hspace*{-0mm}
\includegraphics[width=5.9cm,angle=270]{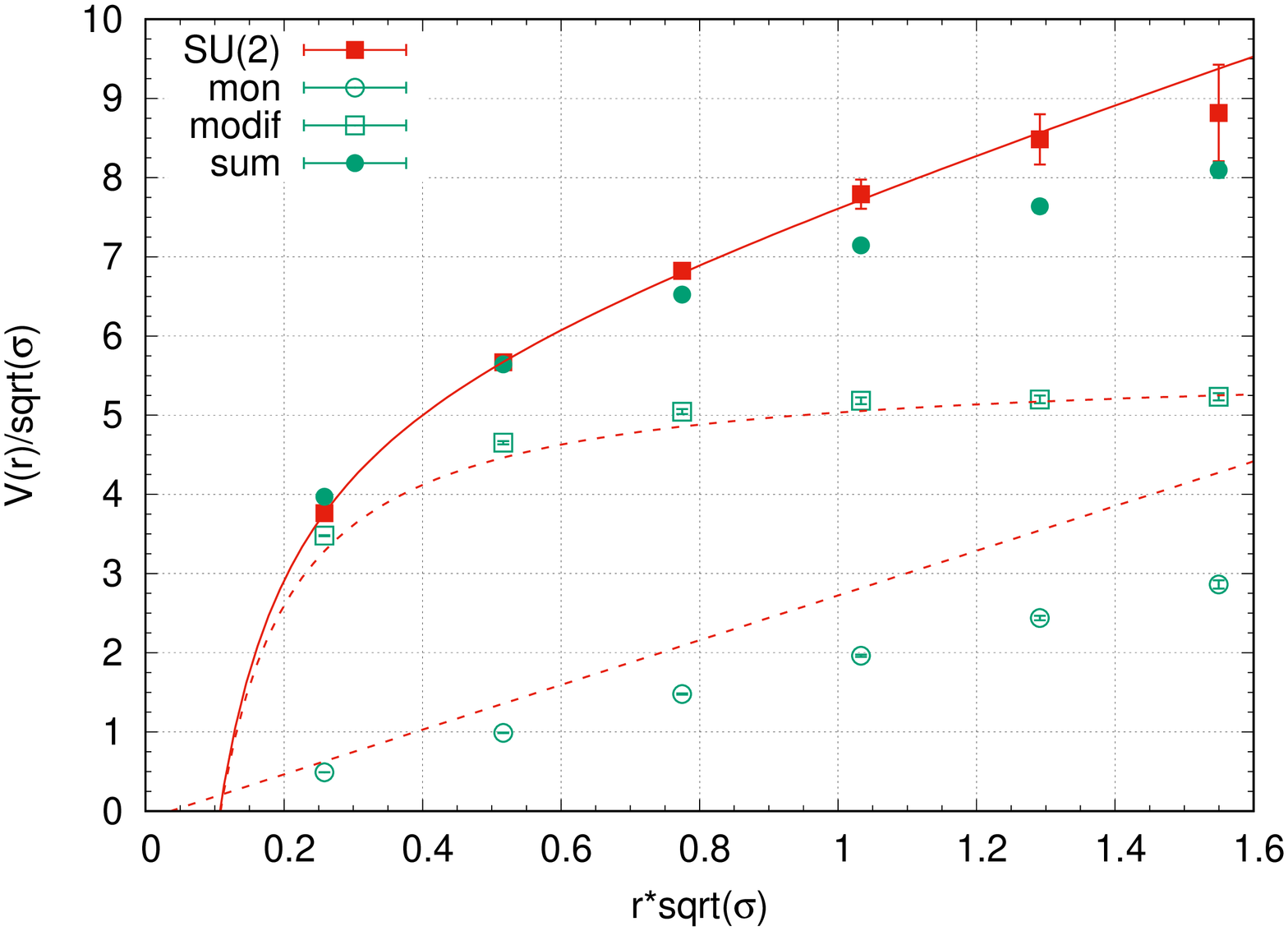}
\includegraphics[width=5.9cm,angle=270]{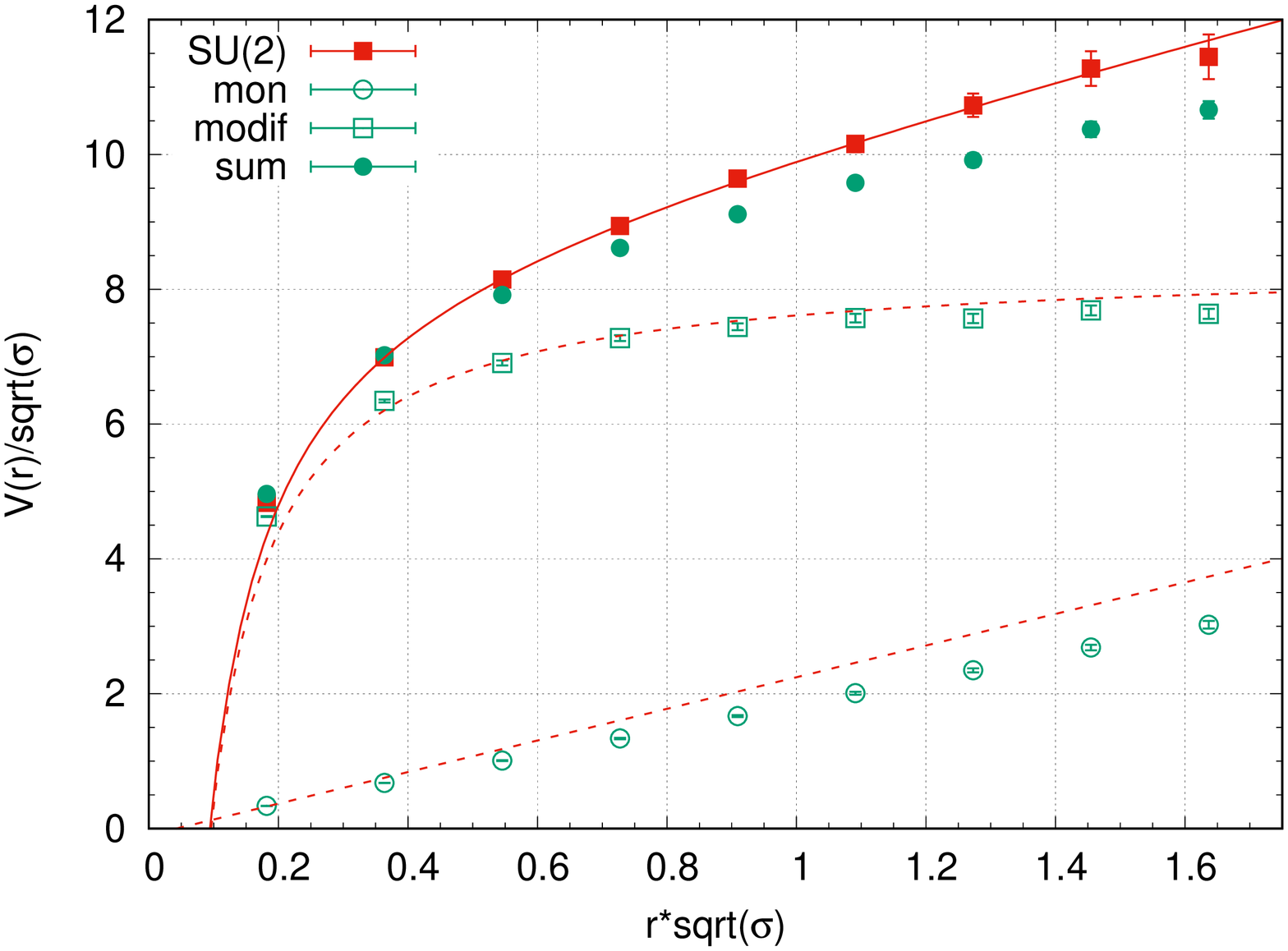}
\hspace*{0mm}
\includegraphics[width=5.9cm,angle=270]{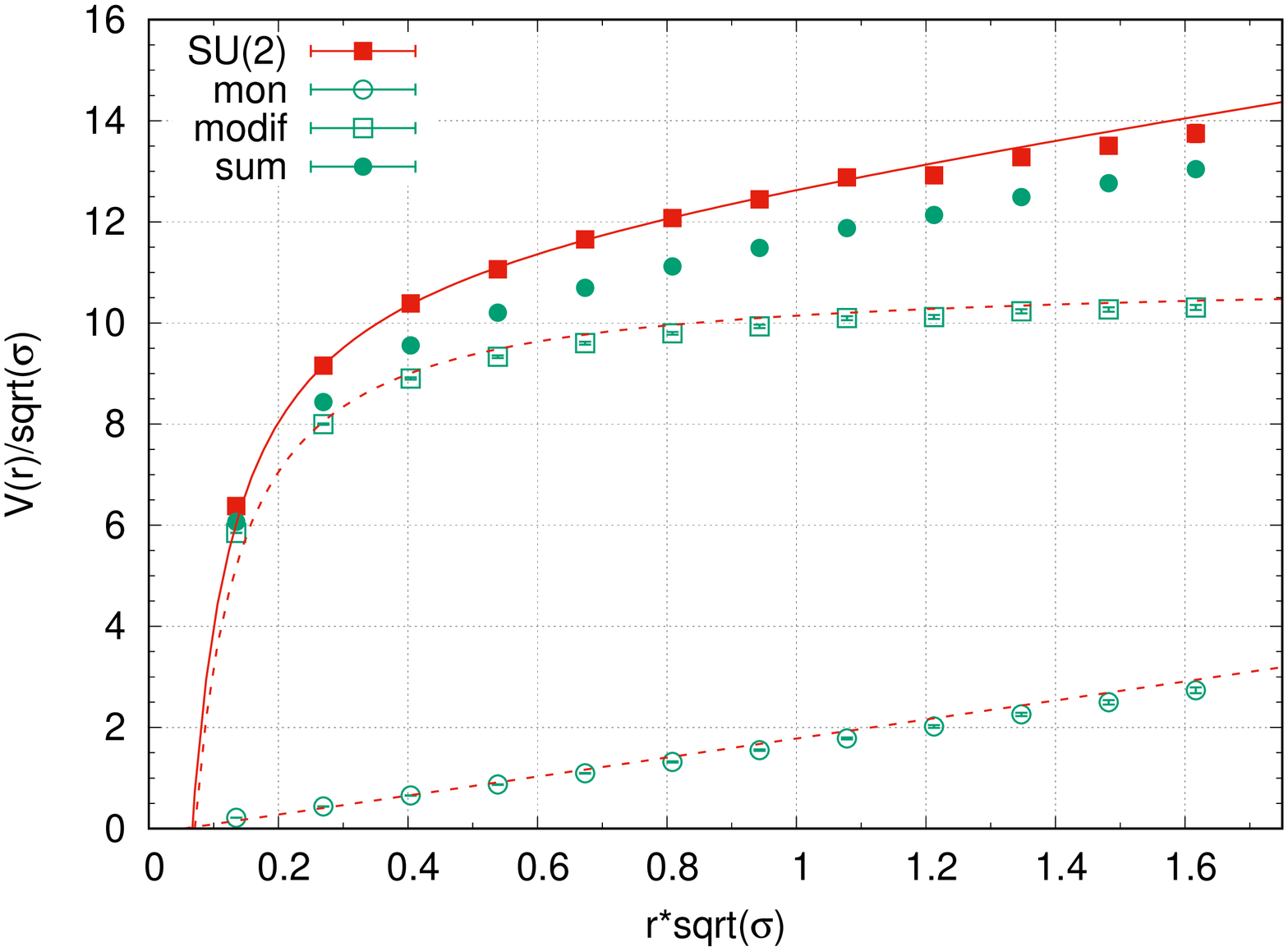}
\includegraphics[width=5.9cm,angle=270]{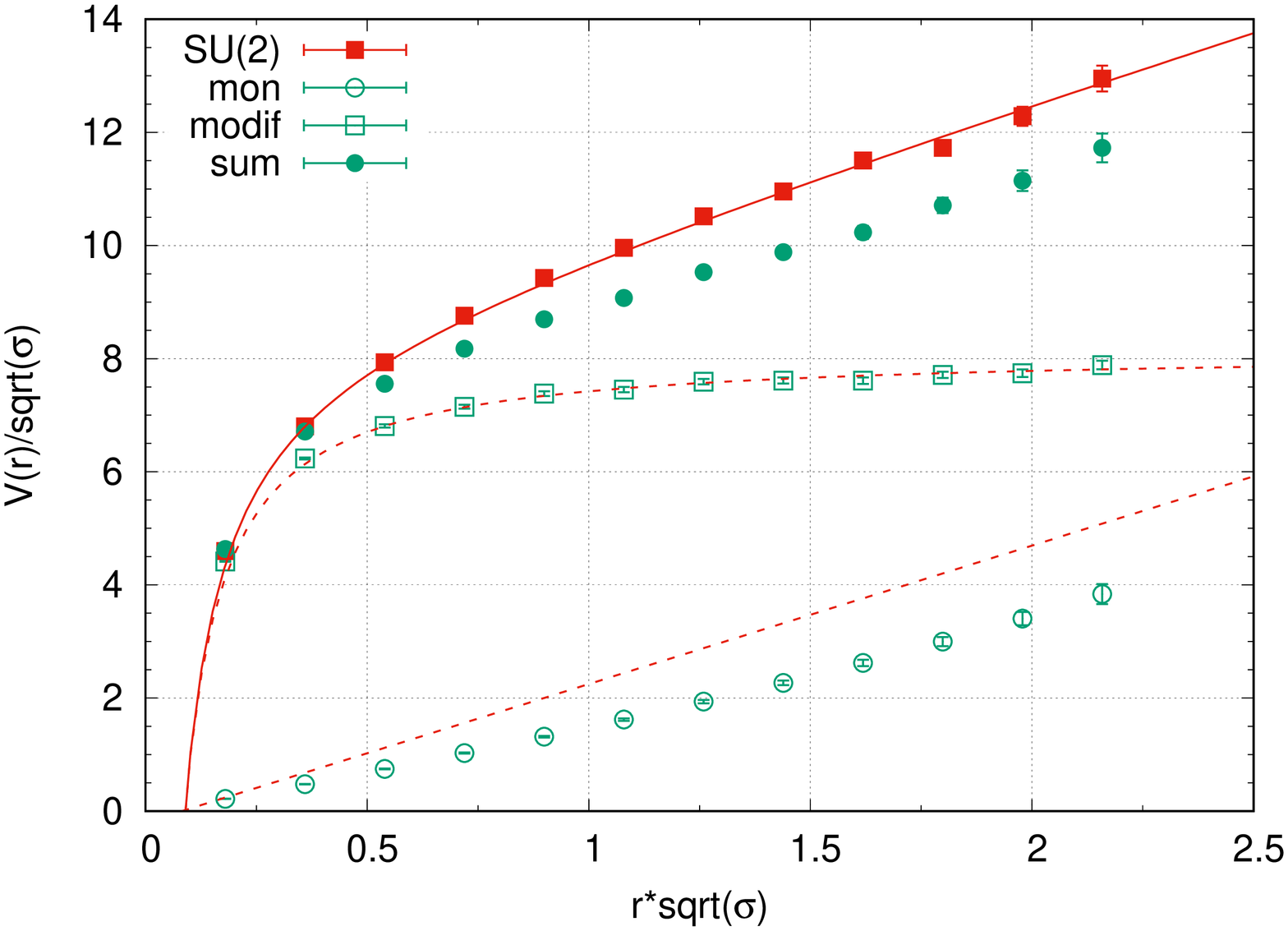}
\caption{Comparison of the adjoint nonabelian potential $V_{adj}(r)$ (filled squares) with the 
sum  $V_{mod,adj}(r)+V_{mon,q2}(r)$ (filled circles) 
for Wilson action at $\beta=2.4$  (left, upper panel),   $\beta=2.5$ (right, upper panel),  $\beta=2.6$  (left, lower panel) and for improved action at $\beta=3.4$ (right, lower panel) . $V_{mod,adj}(r)$ (empty squares) and $V_{mon,q2}(r)$ (empty circles) are also shown. The solid curve and dashed curves carry same meaning as in Fig.~\ref{potentials1}. }
\label{potentials3}
\end{figure}
Furthermore, we did the same study in QC$_2$D on $32^4$ lattice with small lattice spacing (for details of simulations see, e.g. \cite{Bornyakov:2017txe}). The result is presented in Fig.~\ref{potentials2} (right). One can see clearly that approximate decomposition is fulfilled with rather high precision in this case as well.

Next we come to the static potential in the adjoint representation. In this case we check the validity of the relation
\beq
V_{adj}(r) \approx V_{adj,mod}(r) + V_{mon,q2}(r)
\label{eq:adjoint}
\eeq
Our numerical results for three lattice spacing for the Wilson action and for one lattice spacing for the improved action are presented in Fig.~\ref{potentials3}. In this case the precision of our results is lower still it is seen that the relation (\ref{eq:adjoint}) is satisfied quite well.
The signature of improving the agreement between lhs and rhs in (\ref{eq:adjoint}) with decreasing lattice spacing is also seen although this should be checked in more precise measurements.

\section{Conclusions}
We studied the decomposition of the static potential in the fundamental and adjoint representations into the linear term produced by the monopole (Abelian) gauge field $U_{mon}(x)$ and the Coulomb term produced by the monopoleless nonabelian gauge field $U_{mod}(x)$. 
We confirm the results of Ref.~\cite{Bornyakov:2005hf} and improve them in a few respects.
First, we made computations with varying lattice spacing and found that in both representations the agreement becomes better with decreasing lattice spacing. Our results suggest that the relations (\ref{decomp}) and (\ref{eq:adjoint}) become exact in the continuum limit. 
Further work is needed to provide more evidence for this conclusion.
Second, we checked that the decomposition is valid also in the case of improved lattice action and in the theory with quarks. These results make it even more interesting to check this decomposition 
in the case of $SU(3)$ gauge group. 

There are few conclusions to be drawn from the decomposition (\ref{decomp}). It suggests that the monopole part $U_{mon}(x)$ is responsible for the classical part of the hadronic string energy while the monopoleless part $U_{mod}(x)$ produces the fluctuating part of that energy, i.e. while
at small distances $U_{mod}(x)$ should reproduce the perturbative results at large distances it contributes to the nonperturbative physics. 

\acknowledgments{Computer simulations were performed on the 
Central Linux Cluster of the NRC ``Kurchatov Institute'' - IHEP  (Protvino) 
and Linux Cluster of the NRC ``Kurchatov Institute'' - ITEP (Moscow).
This work was supported by the Russian Foundation for Basic Research, 
grant~no.20-02-00737~A. The authors are grateful to G.~Schierholz, T.~Suzuki, S. Syritsyn, V. Braguta, A. Nikolaev for participation at the early stages of this work and for useful discussions.}

\bibliographystyle{plain}
\bibliographystyle{apsrev}
\bibliography{citations_asym_2016}

\end{document}